\pdfoutput=1
\documentclass[preprint,12pt]{elsarticle}

\usepackage{graphicx}
\usepackage{amssymb}
\usepackage{amsmath} 
\usepackage{siunitx} 
\usepackage{xcolor}

\journal{Journal of Magnetic Resonance}

\begin{document}

\begin{frontmatter}


\title{Mapping oscillating magnetic fields around rechargeable batteries}


\fntext[CCL]{Posted with CC BY-NC-ND license}
\author[NYU]{Stefan Benders}
\author[NYU]{Mohaddese Mohammadi}
\author[RIT]{Matthew J. Ganter}
\author[NRL,NYU]{Christopher A. Klug} 
\author[NYU]{Alexej Jerschow\corref{cor1}}
\cortext[cor1]{Corresponding author}
\address[NYU]{Department of Chemistry, New York University, New York, United States}
\address[RIT]{Battery Prototyping Center, Rochester Institute of Technology }

\address[NRL]{United States Naval Research Laboratory, District of Columbia, United States}

\begin{abstract}
Power storage devices such as batteries are a crucial part of modern technology. The development and use of batteries has accelerated in the past decades, yet there are only a few techniques that allow gathering vital information from battery cells in a nonivasive fashion.  A widely used technique to investigate batteries is electrical impedance spectroscopy~(EIS), which provides information on how the impedance of a cell changes as a function of the frequency of applied alternating currents. Building on recent developments of inside-out MRI~(ioMRI), a technique is presented here which produces spatially-resolved maps of the oscillating magnetic fields originating from  the alternating electrical currents distributed within a cell. The technique works by using an MRI pulse sequence synchronized with a gated alternating current applied to the cell terminals. The approach is benchmarked with a current-carrying wire coil, and demonstrated with commercial and prototype lithium-ion cells. Marked changes in the fields are observed for different cell types.
\end{abstract}


\end{frontmatter}


\section*{Introduction}
\label{S:1}
Power storage devices form the basis for portable electronics technology as well as for the development of electric transportation options. The demand for rechargeable batteries will significantly increase in the coming years, yet technology for adequate assessment of commercial-type cells is currently relatively limited. In particular, it is vital to establish detailed noninvasive diagnostics techniques that would allow identifying defects, predicting lifetimes, and determining critical failure mode progressions. 

A commonly used technique to investigate batteries is electrical impedance spectroscopy~(EIS) \cite{galeotti_performance_2015}. EIS is based on applying alternating current~(AC), and recording the complex impedance as a function of current frequency \cite{barsoukov_impedance_2018}. EIS is further widely used to characterize materials as well as biological tissues based on the frequency response of their electrical conductivity and permittivity \cite{estrela_da_silva_classification_2000,glatthaar_efficiency_2007,halter_electrical_2008}. When used with batteries, EIS is often employed to assess information about internal impedance as a function of temperature, depth of discharge, and at different points in the cycling process \cite{singh_fuzzy_2006,eddahech_remaining_2012,pastor-fernandez_identification_2016}. Changes in the frequency response of the internal impedance of an electrochemical energy storage device provide insights into cell performance, including, for example, measures of the uniformity of the material distribution, of the quality of solid electrolyte interphase~(SEI), of the electrode porosity, and of adsorption kinetics \cite{verma_review_2010,jorcin_cpe_2006,deng_electrochemical_2013}. A major limitation of EIS is that it records only the global values for the device, and no spatial resolution is provided. While there exist techniques such as Localized Electrochemical Impedance Spectroscopy \cite{bandarenka_localized_2013} or scanning electrochemical microscopy (AC-SECM)\cite{eckhard_alternating_2008}, these methods require access to the electrode surface and direct electrical contact, which is problematic for realistic or commercial-type cells and many other devices. 
 
 While conventional MRI is not able to investigate the processes within commercial batteries due to the difficulty for radiofrequency~(RF) to penetrate through their metal casings or through the electrode layers, the inside-out MRI~(ioMRI) approach is able to reveal important information by analyzing the magnetic susceptibility changes of the cell's materials via the measurement of the magnetic field distributions around the cell. This approach has been employed to detect susceptibility variations caused by changes in the cell as well as to investigate changes in direct current distributions across the device \cite{romanenko_accurate_2020,mohammadi_diagnosing_2019,mohammadi_situ_2019,ilott_rechargeable_2018,romanenko_distortion-free_2019}. 

In addition, other indirect parameters such as the spatial distribution of the magnetic field generated outside the battery caused by the current can be used as a measure of internal battery properties.

A number of MRI methods have been developed to detect electric currents via measurements of the corresponding magnetic fields using synchronized pulse sequence elements \cite{halpern-manners_magnetic_2010,buracas_imaging_2008,mikac_magnetic_2001,hamamura_measurement_2006,eroglu_induced_2018,kim_detection_2018,yang_mapping_2003}. 
An approach based on a tuned spin-lock field exploits the coherent interaction between the spin-lock pulse and a fluctuating magnetic field produced by electrical currents \cite{halpern-manners_magnetic_2010}. While this approach is potentially quite powerful, there are a number of practical difficulties which limit its successful implementation for our intended application. In particular, this approach exhibits problems at low frequencies due to RF imperfections and $T_\mathrm{1\rho}$ relaxation. Other techniques comprise current density imaging~(CDI) sequences\cite{mikac_magnetic_2001,hamamura_measurement_2006,eroglu_induced_2018} and steady state free precession~(SSFP) approaches\cite{buracas_imaging_2008}, both based on exploiting phase changes in the magnetic resonance signals caused by current-generated local magnetic magnetic fields. However, both approaches require precise synchronization of pulse sequence elements with the AC frequency.

Yang \textit{et al.} showed, that a gradient echo sequence with a sufficient Field-of-View exhibits ghost images at regular distances which depend on the AC frequency of the current passing through a wire at a specific orientation \cite{yang_mapping_2003}. The Fourier transform of the signal revealed these images. More recently, a method based on multiple repetition times to extend the range of frequencies accessible in measurements of brain waves was proposed by Kim \textit{et al.} \cite{kim_detection_2018}. A Fourier transform of a series of images with a fixed repetition time was employed to recover a spectrum of oscillating field frequencies. With this approach, the delay between a certain point of the AC wave and the start of the acquisition is fixed, yet unknown. In this approach, the spectral width of the encoding of the oscillating field is not known. A second experiment set acquired with a different repetition time can recover accurate frequency information without this  \textit{a priori} information.

Building on lessons learnt from prior work, we present here an improved technique, suitable for mapping the oscillating magnetic fields around battery cells. In, particular the use of triggers from the current source to spectrometer as well as gating the AC current allow for more control of the experiment. This sequence and method of potentially performing battery diagnostics is henceforth referred to as inside-out Triggered Alternating Current ImagiNg employing Gating (io-TRACING). In contrast to CDI-based techniques, it is possible to scan a range of frequencies with one measurement instead of having to tailor all pulse sequence element delays to a specific frequency. Furthermore, it is possible to obtain images at very low frequencies, e.g. 0.1\,-\,1\,Hz, which is a particularly useful range  for devices such as batteries. Electric devices, which are the focus here, are excited by an external stimulus, which allows the introduction of gating of the AC, enabling a more controlled experiment. In particular, this method applies a phase-synchronized portion of the AC waveform during the evolution period of the pulse sequence (Figure~\ref{fig:Setup-PP}). In the pulse sequence presented here, a spin echo imaging sequence is used with the evolution period located between the $\frac{\pi}{2}$ and  the $\pi$ pulses. The AC phase is stepped systematically in subsequent measurements, thus producing an indirect time dimension. A Fourier transform along this dimension produces a spectrum containing information about the response of the device to the applied AC voltages. The additional imaging dimensions deliver the spatial localization of these measurements. Results acquired with a range of AC frequencies for a single loop of a wire, a commercial cell and a prototype cell are presented. The comparison between different cells demonstrates that characteristic changes can be observed in such measurements.

\section*{Pulse sequence design and theoretical background}
\label{S:2}

\begin{figure}
    \centering
    \includegraphics[width=\linewidth]{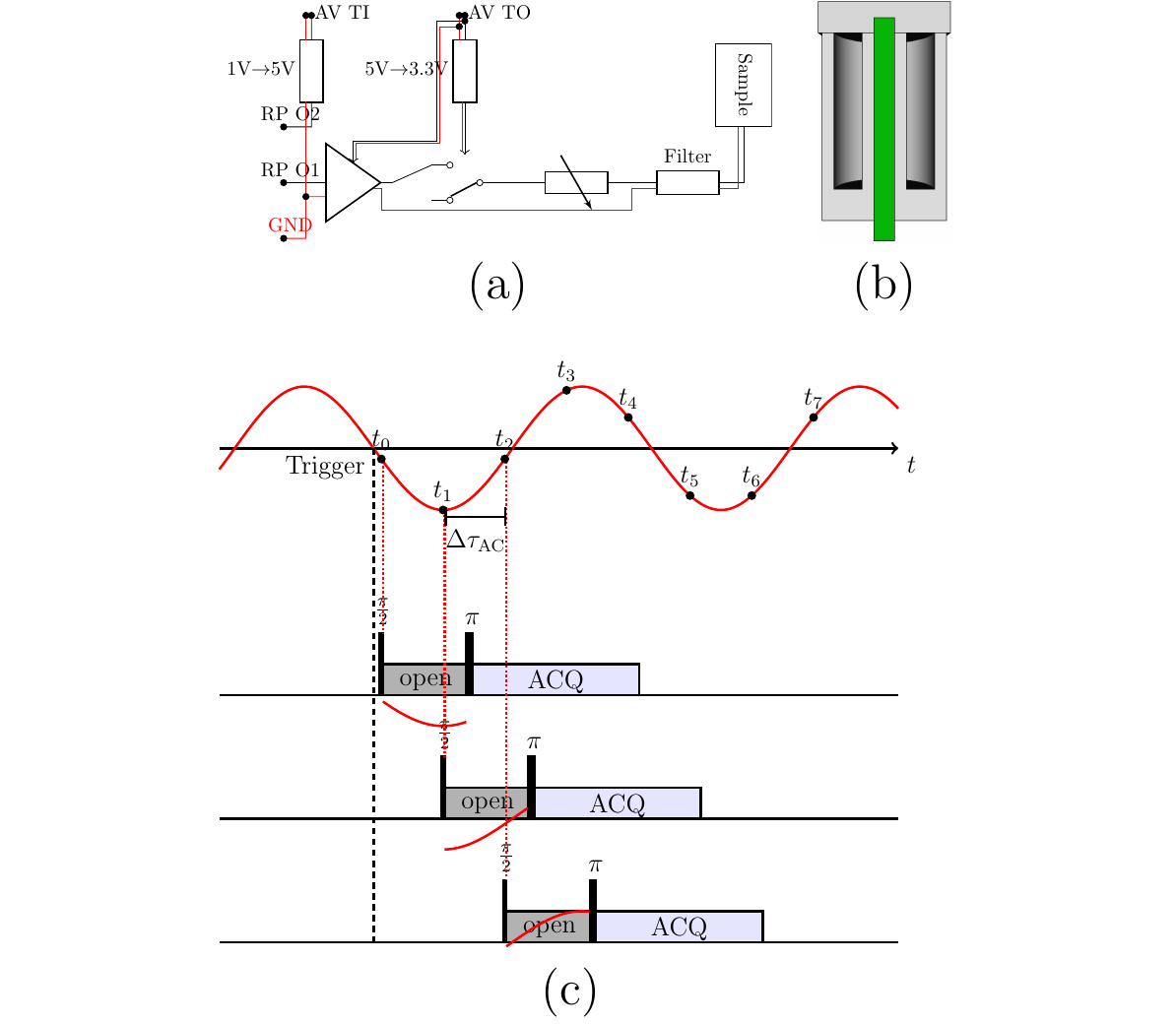}
   \caption{(a) Setup with all electronic components. See Setup and Experimental section for details. (b) 3D-printed holder for the inside-out MRI approach. The battery or coil (green) is located in the middle of the cylindrical holder with two doped water compartments located on each side of the cell/coil. The leads are soldered to the tabs/coil from the bottom of the cell and then guided down the magnet. (c) Pulse sequence. A standard spin echo imaging sequence (Multi-slice-multi-echo) was employed. The sequence was modified to allow for sampling through different delay times after the trigger as well as switching the current on and off after/prior to the pulses. In each encoding segment between the pulses, a different phase is encoded, depending on the starting point $t_n$ (depicted exemplary for point 3), which depends on ${\Delta \tau_\mathrm{AC}}$ with $t_n = t_0 + n \cdot {\Delta \tau_\mathrm{AC}}$. }
    \label{fig:Setup-PP}
\end{figure}

This work employs a spin-echo imaging sequence (Multi-Slice-Multi-Echo, Bruker ParaVision) with modifications to allow for encoding of oscillating magnetic fields. The most significant modification is the insertion of a gate pulse in the interval between the $\frac{\pi}{2}$ and $\pi$ degree pulses which controls the AC current (Figure~\ref{fig:Setup-PP}c). Different portions of the AC wave are sampled in this interval in a systematic way, which is defined by its starting point $t_n$ relative to a well-defined point in the AC wave. Experimentally, this is achieved by employing a synchronized square wave triggering the spectrometer to guarantee a start at the same position in the waveform in consecutive cycles. The total encoding time $\tau$ is held fixed, but the starting time is shifted by a multiple of the sampling delay $\Delta \tau_\mathrm{AC}$ in separate experiments (Figure~\ref{fig:Setup-PP}a,c). This means that the first segment will be sampled with the interval [$0,\tau$], while the next segment is sampled with [$\Delta \tau_\mathrm{AC}$, $\Delta \tau_\mathrm{AC}+\tau$], etc; 16 or 32 of such measurements are typically taken to provide sufficient discrimination between different AC frequencies. The sequence of these experiments allows recording the indirect time dimension along which the AC frequency is encoded. 

The AC signal is encoded in the magnetization evolution as follows. After the excitation pulse to rotate the magnetization into the transverse plane, the magnetization evolves under the influence of the local magnetic field  (Figure~\ref{fig:Setup-PP}c). The field at a given position $\boldsymbol{r}$ in space is composed of a static part and a superimposed oscillating field $B_{z,\mathrm{osc}}$. In a frame rotating at the Larmor frequency of the static field ($\omega_0/2\pi$)  the signal evolution can be described by

\begin{equation}
    S(\boldsymbol{r},t_n)= M_\mathrm{0}(\boldsymbol{r}) \cdot \exp \left\{-i\gamma B_{z,\mathrm{osc.}}(\boldsymbol{r})\cdot \int_{t_n}^{t_n+\tau} \sin(\omega_\mathrm{AC}\cdot t')  \mathrm{d}t'\right\}. \label{eq:1}
\end{equation}

$S(\boldsymbol{r},t_n)$ is the signal encoded by an oscillating field in the interval  $[t_n\, t_n+\tau]$ at a position $\boldsymbol{r}$. $M_\mathrm{0}$ is the bulk magnetization before applying the oscillating field, $\gamma$ is the gyromagnetic ratio and $B_{z,\mathrm{osc}}(\boldsymbol{r})$ and $\omega_{AC}$ are respectively the magnitude and frequency of the oscillating magnetic field. The time $t_n$ indicates the starting point of the AC wave at the beginning of the evolution period and increments with ${\Delta \tau_\mathrm{AC}}$ at each step (as illustrated by the numbered points (0, 1, 2, 3,..) in Figure~\ref{fig:Setup-PP}). One can therefore write  $t_n = t_0 + n \cdot {\Delta \tau_\mathrm{AC}}$, with $t_0$ being  the delay between trigger and AC gating (approximately 6 ms in the experiments done here, the exact delay is not important as long as it is constant).  

To simplify discussions, the location-dependence in Equation~(\ref{eq:1}) and associated symbols will be omitted, with the understanding that location-resolved signals will be provided via the imaging portion of the pulse sequence. 

The integral can be solved and replaced by an infinite sum of cosine terms via the Jacobi-Anger expansion,
\begin{equation}
\begin{split}
S(t_n) & = \left( J_\mathrm{0} \left(\frac{\gamma B_{z,\mathrm{osc.}}}{\omega_\mathrm{AC}}\right) + \sum_{k=1}^{\infty} 2 i^k J_\mathrm{k} \left(\frac{\gamma B_{z,\mathrm{osc.}}}{\omega_\mathrm{AC}}\right) \cos \left(k \cdot \omega_\mathrm{AC}\cdot t_n + k \cdot \omega_\mathrm{AC}\cdot \tau \right) \right) \\
\cdot & \left( J_\mathrm{0} \left(-\frac{\gamma B_{z,\mathrm{osc.}}}{\omega_\mathrm{AC}}\right) + \sum_{l=1}^{\infty} 2 i^l J_\mathrm{l} \left(-\frac{\gamma B_{z,\mathrm{osc.}}}{\omega_\mathrm{AC}}\right) \cos \left(l \cdot \omega_\mathrm{AC}\cdot t_n \right) \right)
\end{split}. \label{eq:jaexp}
\end{equation}

\begin{figure}
	\centering
	\includegraphics[width=0.75\linewidth]{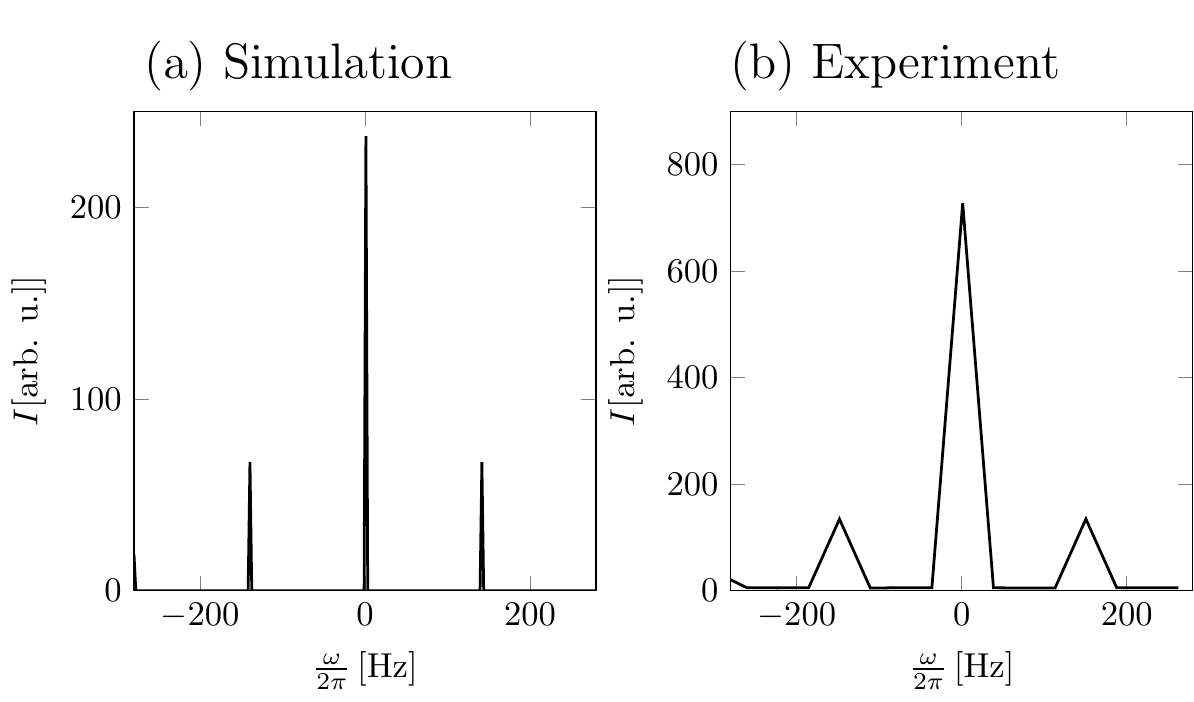}
	
	\caption{Left: Simulation of the io-TRACING spectrum at 140\,Hz AC frequency and 10\,ms echo time with 1024 points of AC sampling and a $\Delta \tau_\mathrm{AC}$ of 1.786\,ms. Right: Result from an experiment with 16 points of AC sampling. The intensities of all pixels in the frequency domain are shown for an experiment with $\omega_{AC}/2\pi$=140\,Hz AC frequency and 54\,mA AC current.}
	\label{fig:compspec}
\end{figure}

As can be seen, a series of oscillating terms at multiples of the base frequency arise. Further details of the derivation are provided in the Appendix. While the series expansion provides conceptual insights, a numerical calculation of  Eq. (\ref{eq:1}) and its Fourier transform was found more practical, and was performed using MatLab (MathWorks, Natick, MA). These simulations obtained for a single frequency and fixed $\tau$ show the expected behaviour of peaks at multiples of the AC frequency and zero as seen in Figure~\ref{fig:compspec}. Plotting the ratio of the respective frequency peak to the zero-frequency peak versus AC frequency reveals a magnitude sinc-shaped behaviour, i.e. a general decay with AC frequency as well as a substructure leading to zero intensity nodes at multiples of $\frac{1}{\tau}$ (Figure~S1). A choice of a small $\tau$ changes the position of these nodes, but reduces the encoding strength. 

The strong dependence on $\frac{\omega_\mathrm{AC}}{2\pi}$, with the envelope in Figure~S1 decaying with a multiple of $\frac{1}{\frac{\omega_\mathrm{AC}}{2\pi}}$, also suggests that this method allows the measurement of frequencies of up to approximately 1-2\,kHz. 

The spectral width for AC-frequency encoding is given by SW$_\mathrm{AC} = \frac{1}{\Delta \tau_\mathrm{AC}}$ (Figure~\ref{fig:Setup-PP}). As a consequence, this method can be employed to scan for an unknown frequency by employing a small sampling delay and many steps. The acquisition can be tailored to a specific frequency range by adjusting the sampling delay according to this expression. A compromise can often be found between sufficient spectral resolution and experiment duration. This approach is particularly important at low frequencies. For example, for 0.1\,Hz, the sampling delay becomes 2.5\,s for a io-TRACING spectral width of 0.4\,Hz, and the experiment times for 8, 16 or 32 points in the indirect AC dimension (phase dimension 64 points, 4 scans, TR 125\,ms) become roughly 6.4\,h, 24.2\,h and 93.9\,h, respectively. The inconvenient lengthening of experiment times at very small frequencies is also known in EIS. 

The spectral resolution is given by the spectral width for AC-frequency encoding and the number of encoded segments $n_\mathrm{enc}$: $\frac{\mathrm{SW}_\mathrm{AC}}{n_\mathrm{enc}} = \frac{1}{\Delta \tau_\mathrm{AC} \cdot n_\mathrm{enc}}$.The sampling delay should be set to capture all frequencies in the sample, leading to the conditions $\mathrm{SW}_\mathrm{AC} > 2\frac{\omega_\mathrm{AC}}{2\pi}$, so $\Delta \tau_\mathrm{AC} < \frac{\pi}{\omega_\mathrm{AC}}$, and $n_\mathrm{enc}$ set to distinguish between different frequencies, or the frequency and the zero peak, so $\frac{\mathrm{SW}_\mathrm{AC}}{n_\mathrm{enc}} <  \frac{\Delta \omega_\mathrm{AC}}{2\pi} \rightarrow n_\mathrm{enc} > \frac{2\pi\mathrm{SW}_\mathrm{AC}}{\Delta \omega_\mathrm{AC}} \rightarrow n_\mathrm{enc} > \frac{2\pi}{\Delta \omega_\mathrm{AC} \Delta \tau_\mathrm{AC}}$. The encoded signal (after Fourier transformation along this dimension) is proportional to the field in the low-field regime for a specific $\tau$ (for $\tau$=3.1\,ms and $B<$1.5\,\si{\micro}T) (Figure~S2), hence it is a direct measure of the $z$ field strength at a certain frequency. There is a functional relationship between intensity and frequency (Figure~S1) , which requires precise calibration with the encoding time $\tau$. While this parameter can be set, switching times of electronic components alter the value. Here, a field estimate based on simulations is pursued, which is inaccurate at frequencies close to $\frac{1}{\tau}$. In principle, experimental calibration is possible with e.g. a coil fixed in position and immersed in the medium.

\section*{Setup and experimental}

The experimental schematic is shown in Figure~\ref{fig:Setup-PP}a. The AC waveform was generated with a Red Pitaya STEMLab 125-14 FPGA controller unit. The generated signal was amplified with a Stanford Research Systems SR560 Low Noise-Amplifier (Gain 1000). To trigger the spectrometer from the Red Pitaya control unit, a square wave was generated on the second output of the unit and then shifted from 1\,V to 5\,V employing an Arduino UNO. The amplifier blanking was realized with an output of the spectrometer. In the case of batteries, a switch (PE42020, Peregrine) was employed to prevent DC discharge against the output impedance of the amplifier when AC current is not applied to the battery. This switch is able to reliably disconnect the battery from the generation at the frequencies employed in this study. The output is fed to the measurement cell through a variable resistor to fine tune the current level, and a filter to remove high frequency components in the RF region.

The experiment was set up such that the magnetic fields were mapped around the battery with a suitable detection medium and MRI sequence according to the ioMRI strategy \cite{ilott_rechargeable_2018,mohammadi_diagnosing_2019}. A polylactic acid~(PLA) cell holder (outer diameter 19.75\,mm) was 3D-printed consisting of a slot to hold the battery and two semicircular chambers next to that slot (Figure~\ref{fig:Setup-PP}b). The chambers were filled with aqueous 0.1\,M CuSO$_\mathrm{4}$ solution. The coil used in the evaluation experiments consisted of a single loop with a diameter of 18\,mm. Two batteries were employed in this study. (1) A Rochester Institute of Technology (RIT) cell: Pouch cell Li-ion battery with 250\,mAh capacity. The battery is made of five layers of stacked-cut graphite anode and NMC cathode electrodes. The cathode is composed of Co(5.97), O(19.46), Ni(14.46), and Mn(9.19). (2) Powerstream (PS) cells: Jelly rolled lithium ions battery with 600\,mAh capacity. The graphite and NMC electrodes are rolled in twelve active layers and packed inside an aluminum pouch case. Both batteries are made from graphite anode, aluminum and copper current collectors. The cathode in the PS battery is made of Co (44.76), O (33.20), Ni (4.79), Mn(2.99). Both cells were fully charged prior to the experiment. The dimensions of the RIT and PS cells are 65.7\,$\times$\,42.5\,mm$^2$, and 40\,$\times$\,30\,mm$^2$, respectively. Consequently, the battery tabs of the RIT cell are located further out of the measurement volume than the PS cells. Since the holder has a 10\,mm bottom to prevent leaking, the lowest 10\,mm of the battery close to the tabs are not represented. Furthermore, on the opposite side, the image is limited by the excitation volume of the RF coil.

A 400\,MHz Bruker Avance magnet with a 57\,mm gradient system (max. gradient strength 0.3\,T/m) and a 40\,mm $^1$H coil was employed for the experiments. The battery or coil was rotated with the holder to align parallel with the $B_\mathrm{1}$ field to minimize artifacts. The main experimental parameters were set as follows: echo time 10\,ms (except stated otherwise); repetition time 120\,ms; number of averages 2; number of repetitions (AC frequency dimension, number of different phases sampled) 8 (PS cell\,$<$\,0.25\,Hz)/16 (PS cell\,$>$\,0.25\,Hz, coil)/32 (RIT cell); sampling delay for the AC frequency: $\frac{1}{4\cdot\frac{\omega_\mathrm{AC}}{2\pi}}$ (set with the expected AC frequency, see the section on pulse sequence design and theoretical background; excitation pulse length 2\,ms  (90$^\circ$); refocusing pulse length 1.267\,ms , field-of-view (except different orientations) 40\,$\times$\,30\,mm$^2$; matrix size 128\,$\times$\,64/32; slice thickness 2\,mm; dwell time 20\,\si{\micro}s. 

The number of samples in the AC dimension is NR, and the increment of the delay $t_n$ is ${\Delta \tau_\mathrm{AC}}$. For each of these, a full $k$-space is acquired. 

For a two-dimensional image such as those obtained in this study, the data is subject to a three-dimensional Fourier transform, where two dimensions represent image dimensions and one dimension the AC field frequency dimension. Each pixel of the image is then associated with a spectrum of NR points. To identify the dominating frequencies, all pixels are integrated and the three highest peaks are identified. Then, the data point number of e.g. the positive peak is identified, its intensity set in relation to the zero-frequency peak intensity, and their ratio is plotted as an image.

\section*{Evaluation of the technique with a single loop coil}

A single-loop coil was used to test the measurement setup and to demonstrate the ability to map the oscillating magnetic fields. According to the presented theoretical considerations, the Fourier transform of the signal features peaks at zero and $\pm n \frac{\omega_\mathrm{AC}}{2\pi}$. If the corresponding spectra of all pixels are summed up, a global spectrum at a frequency of $\frac{\omega_\mathrm{AC}}{2\pi}$ is obtained, which shows the expected behavior (Figure~\ref{fig:compspec}, right). Overtone signal peaks may be observed at lower intensities at multiples of the base frequency. 

\begin{figure}
    \centering
    \includegraphics[width=\linewidth]{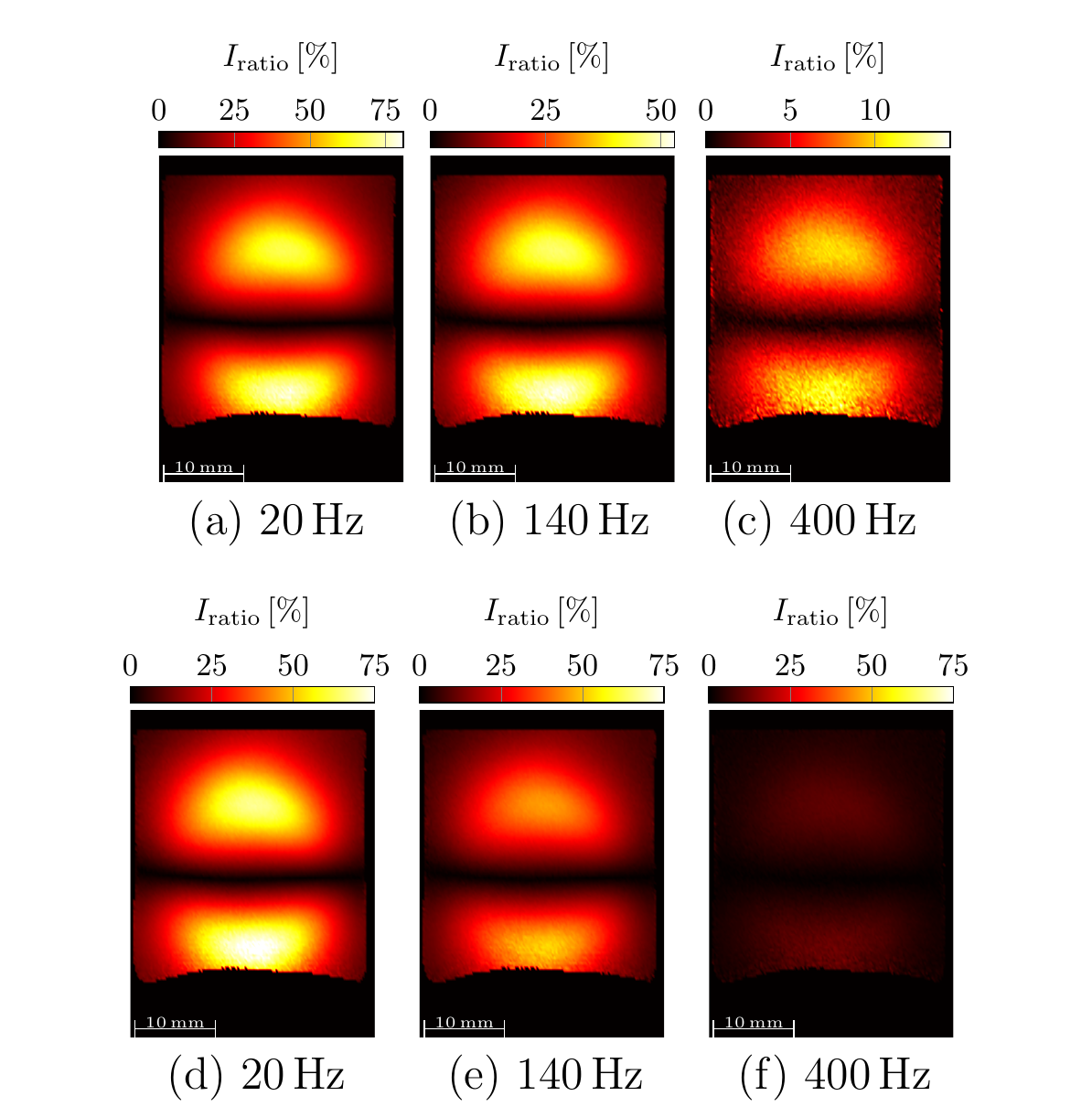}
    \caption{Comparison of the positive frequency part of the spectrum at different AC frequencies with a current of 54\,mA. While the absolute intensity changes (bottom row), the relative patterns remain (top row). 16 experiments with different parts of the waveform have been employed for reconstruction resulting in $t_n$ of 0 to 15~$\Delta \tau_\mathrm{AC}$.}
    \label{fig:CompACfreq}
\end{figure}

The spatial distribution of the magnetic field does not change with frequency for the single loop measurements. This is expected as the coil impedance does not change in the frequency range of the applied current (0.1\,-\,4000\,Hz). In the images of a slice parallel to the coil plane, the $z$-component of the field can be observed with a zero node in the middle of the coil and two regions of high intensity (Figure~\ref{fig:CompACfreq}). A plane parallel to the coil at a certain distance is expected to show a $z$-field, which  increases from the center position upwards   and then decreases beyond the coil radius (see the simulations in Figure~S3). The sign of the field was not determined with this approach due to magnitude processing. The phase of single pixels is frequency dependent, as the Fourier terms in the supporting information reveal, but the dependence is complex. In the future, it may be possible to obtain phase information of the field by employing a phase-sensitive acquisition scheme.

\begin{figure}
    \centering
    \includegraphics[width=0.75\linewidth]{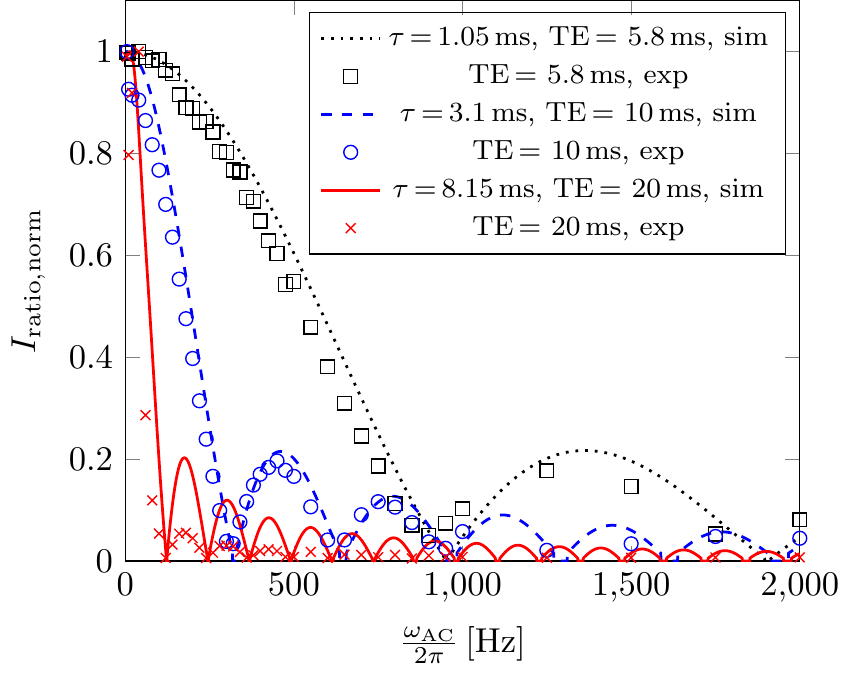}
    \caption{Comparison of experimental (positive frequency, single loop) and simulated data. The experimental data represent an integral of the whole image and all curves have been normalized for comparison purposes. The encoding time for the AC frequency, $\tau$ is related to the echo time (TE) via  $\tau = \frac{\mathrm{TE}}{2}-1.85\,\mathrm{ms}$, since the AC field is turned on only during the first echo period, and TE also includes additional pulse durations. 16 experiments with different parts of the waveform have been employed for reconstruction of each point.}
    \label{fig:ExpSimComp}
\end{figure}

While the pattern of the observed field magnitude remains largely the same as a function of frequency, the overall intensity changes, as expected from numerical simulations of Equation~\ref{eq:1} (Figure~S1). Furthermore, the encoding strength can be controlled by a change of the echo time, or more precisely the encoding time $\tau$ (Figure~S4).  In the bottom row of Figure~\ref{fig:CompACfreq}, the intensity variation with frequency can be observed. To compare field pattern changes, it is desirable to have a common scale. Therefore, the images have to be multiplied by a factor related to the physical field, which can be estimated from simulations. The integrated signal ratio between the first frequency peak at $\omega=\omega_\mathrm{AC}$ and the $\omega=0$ peak over the whole image at these frequencies should therefore follow the simulated response. In Figure~\ref{fig:ExpSimComp}, the experimental data agree relatively well with the simulation, although the encoding time seems to be slightly different due to electronic part switching times, causing shifts in the pattern. Furthermore, the intensity of the 20\,ms echo time curves do not seem to line up. At such a high echo, the encoding of field information is strong, which causes inverted ratios of $\omega=0$ and $\omega=\omega_\mathrm{AC}$ peaks and therefore elevated errors. This is enhanced by the normalization of the intensity in this Figure in order to compare between simulation and experiment. Smaller encoding/echo times lead to lower intensity as also shown in Figure~S1. Furthermore, as expected, the intensity is a function of current and scales proportionally (Figure~S5). 

The numerical simulations can also serve as a way to estimate the field. The ratio of peaks, which has a magnitude sinc-like behaviour with AC frequency as shown in Figures~\ref{fig:ExpSimComp} and S1, can be employed to calculate a factor to transform the data into field estimates. However, due to small differences in encoding time caused by electronic component switching, the curves of experiment and simulation do not completely agree with each other and seem to be shifted. Therefore, especially at frequencies close to the zero-intensity nodes, this method is not very accurate. The peak field amplitude is estimated to be 1.5\,\si{\micro}T. At half of the current, the intensity is also approximately halved leading to a field estimate of 0.75\si{\micro}T (Figure~S5). A simulation with in-house Matlab code shows the same field pattern and reproduces the observed field values up to a residual discrepancy at a level of 0.93\,\si{\micro}T (Figure~S3, 7.5\,mm slice position). Small differences in sample positioning lead to strong changes in magnitude. Therefore, a small offset of the distance leads to a slightly larger or smaller value. Evidence for this assumption can be found in Figure~S6~(b), where both sides show different intensities.

\section*{Application of the technique to commercial batteries}
\begin{figure}
    \centering
    \includegraphics[width=0.75\linewidth]{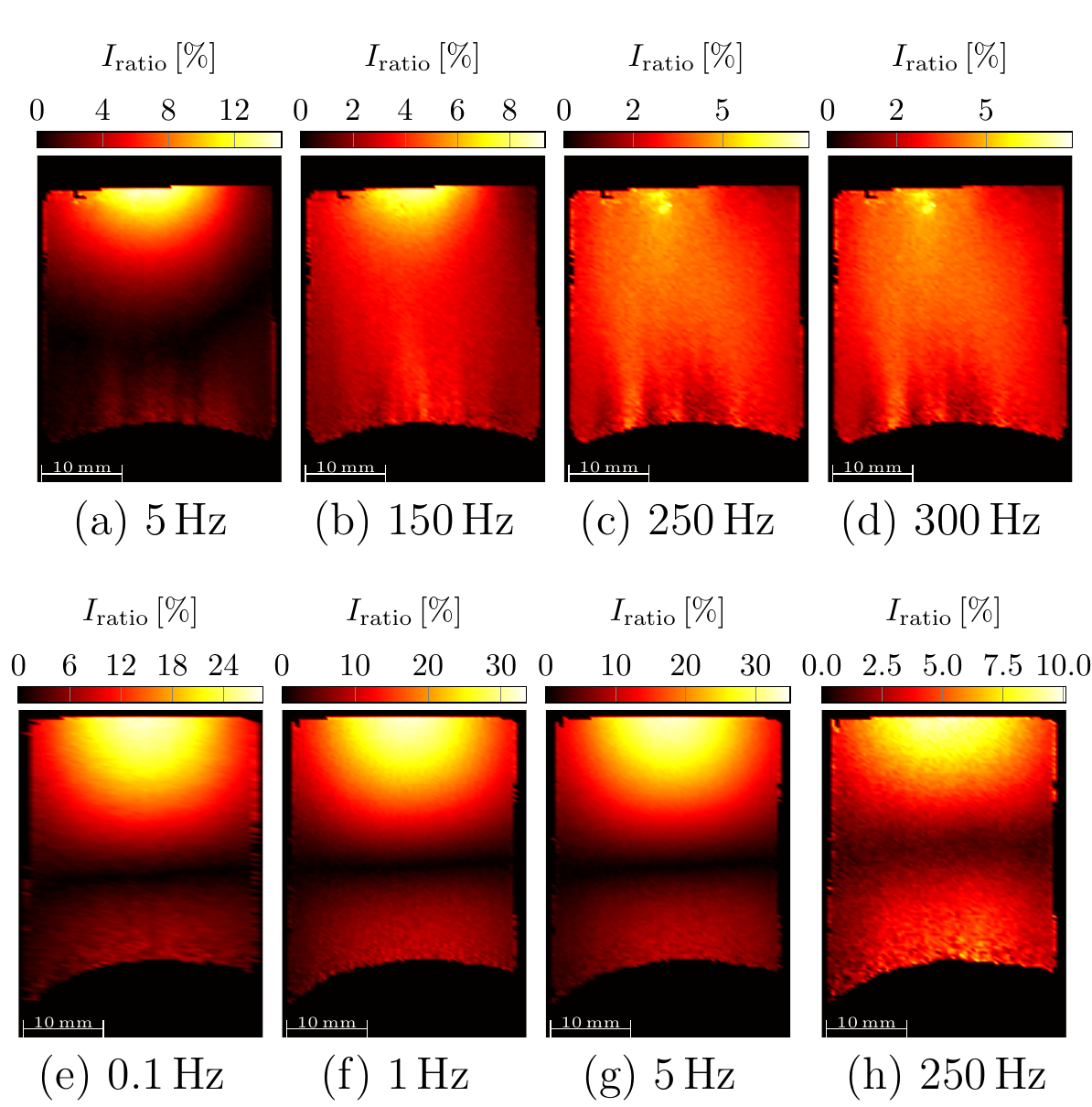}
    \caption{Relative measurements of the oscillating fields generated by 50\,mA AC current in various Li-ion batteries at different driving AC frequencies. (a)-(d) are maps from a 250\,mAh RIT cell, which has a  stacked electrode design. The images in (e)-(h) are from a 600\,mAh PS jelly-rolled cell. 16 experiments with different parts of the waveform have been employed for reconstruction of each point of the PS images ($t_n$ from 0 to 15~$\Delta \tau_\mathrm{AC}$), while 32 were used for the RIT cell ($t_n$ from 0 to 31~$\Delta \tau_\mathrm{AC}$). A different scaled version can be found in Figure~S7}
    \label{fig:BattACfreq}
\end{figure}

 The battery tabs are located at the top of the images shown in Figure~\ref{fig:BattACfreq}. Measurements of a PS jelly-rolled 600\,mAh cell show small changes with frequency. A zero node can be observed at all frequencies (Figure~\ref{fig:BattACfreq}~(e-h)). One notable observation is that the thickness of the zero node in the images decreases slightly with frequency. Furthermore, a small change of the top pattern can be observed at low frequencies. Another change observed at very low AC frequencies (below 2\,Hz) is a broadening of the io-TRACING spectrum peaks to more than one point, leading to minor intensity reductions in the corresponding images. The peak field estimate for this cell is in the range of 700\,nT. 
 
 For a RIT cell with 250\,mAh capacity, there are significant changes with frequency. While at lower frequencies, the intensity is at the tabs (Figure~\ref{fig:BattACfreq}~a), the pattern changes significantly with frequency (Figure~\ref{fig:BattACfreq}~b-d). The field is in the range of 300-1500\,nT.
 
 These results are encouraging with respect to demonstrating the ability to detect AC fields originating from cells. The fact that marked differences between the different types of cells are seen shows that battery characteristics do leave their marks on the measurements. The two cell types differ in their designs, capacities, and form factors. Notably one has stacked electrodes, and the other rolled ones. Furthermore, the observation that the pattern changes with frequency for only one of them indicates that the AC measurement holds a lot more information than can presently be extracted. 
 
 It is recognized, however, that the present work does not constitute proof that the AC measurement provides significant improvements in the diagnostic ability or differentiation between cells. More work with a careful selection of cell types, designs, and properties, would be necessary to delineate the connection between underlying changes in electrochemical behavior and the measured oscillating fields in cells, which  exceeds the scope of this article. 
 
  Future improvements of the method could include acceleration with advanced imaging schemes such as Rapid Acquisition with Relaxation Enahancement (RARE)\cite{hennig_rare_1986}, a constant-time implementation to remove the frequency dependence on the signal intensity, and a lock-in detection to extract the sign and phase of the field and improve the quantification of the field. The developed methodology could also be applied to devices other than the  battery cells considered here, including, for example, fuel cells, and other electrochemical testing chambers in both inside-out and conventional implementations. 
 
\section*{Conclusions}
A new technique is presented, allowing measuring  oscillating magnetic fields around battery cells. In contrast to many other techniques requiring synchronization of the AC frequency with the whole pulse sequence, in the current io-TRACING approach, the AC sampling conditions can be set independently of the sequence.  The technique relies on triggering from a frequency generator as its only input to insure a consistent start point for the sequence. The method is relatively easy to implement with minor electronic parts external to the spectrometer. In combination with the inside-out MRI (ioMRI) approach, complicated systems such as batteries and other devices can be investigated. In this work, changes between different Li-ion battery types were observed.  The developed technique could be applied to other devices and processes, and may aid in nondestructive battery characterization, by providing information akin to a localized version of electrochemical impedance spectroscopy.

\section*{Acknowledgements}
This work was supported the US National Science Foundation under award CBET 1804723. Furthermore, the authors like to thank Leeor Alon for fruitful discussions. CAK acknowledges support for this research from the Office of Naval Research (ONR) through the base program NRL core funding.

\appendix
\section{}
 \subsection*{Approximations to the Fourier transform of the signal equation}
 
 The integrated signal equation is (with $f_\mathrm{AC}=\frac{\omega_\mathrm{AC}}{2\pi}$):
 \begin{equation}
 S(t_n)= M_\mathrm{0} \cdot \exp \left\{i \cdot \gamma \frac{B_{z,\mathrm{osc.}}}{2\pi f_\mathrm{AC}}\cdot \left( \cos(2\pi f_\mathrm{AC}\cdot (t_n+\tau)) -\cos(2\pi f_\mathrm{AC}\cdot t_n) \right) \right\} 
 \end{equation}
 
 This expression can be rewritten with the help of the Jacobi-Anger expansion
 \begin{equation}
 \exp (iz\cos \Phi) = J_\mathrm{0} (z) + 2 \cdot \sum_{n=1}^\infty i^n J_n (z) \cos(n\Phi),
 \end{equation}
 which leads to:
 
 \begin{equation}
 \begin{split}
 S(t_n) & = \left( J_\mathrm{0} \left(\frac{\gamma B_{z,\mathrm{osc.}}}{2\pi f_\mathrm{AC}}\right) + \sum_{k=1}^{\infty} 2 i^k J_\mathrm{k} \left(\frac{\gamma B_{z,\mathrm{osc.}}}{2\pi f_\mathrm{AC}}\right) \cos \left(k \cdot 2\pi f_\mathrm{AC}\cdot t_n + k \cdot 2\pi f_\mathrm{AC}\cdot \tau \right) \right) \\
 \cdot & \left( J_\mathrm{0} \left(-\frac{\gamma B_{z,\mathrm{osc.}}}{2\pi f_\mathrm{AC}}\right) + \sum_{l=1}^{\infty} 2 i^l J_\mathrm{l} \left(-\frac{\gamma B_{z,\mathrm{osc.}}}{2\pi f_\mathrm{AC}}\right) \cos \left(l \cdot 2\pi f_\mathrm{AC}\cdot t_n \right) \right)
 \end{split}
 \end{equation}
 
 The Jacobi-Anger expansion consists of a series of Bessel functions. Replacing the cosine functions with sums of exponentials using the Euler formula allows formally writing down the Fourier transformation of this expression. As an example, solutions are provided considering cutoffs of $n\le 5$ and $m\le 5$. These expressions are provided  for the zero-frequency peak $I(0)$, and the $\pm f_{AC}$ frequency peak amplitudes $I(\pm f_{\mathrm{AC}})$ are provided below, and a full numerical solution is plotted in Figure~S1.
 
 \begin{multline}
 I(0)=\sqrt{2\pi}  \{\\
 e^{-5i\tau 2\pi f_\mathrm{AC}} J_0^2\left(\frac{\gamma B_{z,\mathrm {osc.}}}{2\pi f_\mathrm{AC}}\right) + \left( e^{4i\tau 2\pi f_\mathrm{AC}}+e^{6i\tau 2\pi f_\mathrm{AC}} \right) J_1^2\left(\frac{\gamma B_{z,\mathrm {osc.}}}{2\pi f_\mathrm{AC}}\right) \\
 + e^{3i\tau 2\pi f_\mathrm{AC}} J_2^2\left(\frac{\gamma B_{z,\mathrm {osc.}}}{2\pi f_\mathrm{AC}}\right) + e^{7i\tau 2\pi f_\mathrm{AC}} J_2^2\left(\frac{\gamma B_{z,\mathrm {osc.}}}{2\pi f_\mathrm{AC}}\right) \\
 + e^{2i\tau 2\pi f_\mathrm{AC}} J_3^2\left(\frac{\gamma B_{z,\mathrm {osc.}}}{2\pi f_\mathrm{AC}}\right) + e^{8i\tau 2\pi f_\mathrm{AC}} J_3^2\left(\frac{\gamma B_{z,\mathrm {osc.}}}{2\pi f_\mathrm{AC}}\right) \\
 +  e^{i\tau 2\pi f_\mathrm{AC}} J_4^2\left(\frac{\gamma B_{z,\mathrm {osc.}}}{2\pi f_\mathrm{AC}}\right) + e^{9i\tau 2\pi f_\mathrm{AC}} J_4^2\left(\frac{\gamma B_{z,\mathrm {osc.}}}{2\pi f_\mathrm{AC}}\right)\\
 +  e^{10i\tau 2\pi f_\mathrm{AC}} J_5^2\left(\frac{\gamma B_{z,\mathrm {osc.}}}{2\pi f_\mathrm{AC}}\right) + J_5^2\left(\frac{\gamma B_{z,\mathrm {osc.}}}{2\pi f_\mathrm{AC}}\right) \\
 \}
 \end{multline}
 
 \begin{multline}
 I(f_{\mathrm{AC}})=i \sqrt{2 \pi } e^{-4i\tau 2\pi f_\mathrm{AC}}  \{ \\
 -1 + e^{i\tau 2\pi f_\mathrm{AC}} J_0\left(\frac{\gamma B_{z,\mathrm {osc.}}}{2\pi f_\mathrm{AC}}\right) J_1\left(\frac{\gamma B_{z,\mathrm {osc.}}}{2\pi f_\mathrm{AC}}\right)\\
 - e^{3i\tau 2\pi f_\mathrm{AC}} \left( -1 + e^{3i\tau 2\pi f_\mathrm{AC}} \right) J_2\left(\frac{\gamma B_{z,\mathrm {osc.}}}{2\pi f_\mathrm{AC}}\right) J_1\left(\frac{\gamma B_{z,\mathrm {osc.}}}{2\pi f_\mathrm{AC}}\right) \\
 + e^{2i\tau 2\pi f_\mathrm{AC}}  J_2\left(\frac{\gamma B_{z,\mathrm {osc.}}}{2\pi f_\mathrm{AC}}\right) J_3\left(\frac{\gamma B_{z,\mathrm {osc.}}}{2\pi f_\mathrm{AC}}\right) - e^{7i\tau 2\pi f_\mathrm{AC}}  J_2\left(\frac{\gamma B_{z,\mathrm {osc.}}}{2\pi f_\mathrm{AC}}\right) J_3\left(\frac{\gamma B_{z,\mathrm {osc.}}}{2\pi f_\mathrm{AC}}\right) \\
 +e^{i\tau 2\pi f_\mathrm{AC}} J_3\left(\frac{\gamma B_{z,\mathrm {osc.}}}{2\pi f_\mathrm{AC}}\right) J_4\left(\frac{\gamma B_{z,\mathrm {osc.}}}{2\pi f_\mathrm{AC}}\right) - e^{8i\tau 2\pi f_\mathrm{AC}} J_3\left(\frac{\gamma B_{z,\mathrm {osc.}}}{2\pi f_\mathrm{AC}}\right) J_4\left(\frac{\gamma B_{z,\mathrm {osc.}}}{2\pi f_\mathrm{AC}}\right) \\
 - e^{9i\tau 2\pi f_\mathrm{AC}} J_4\left(\frac{\gamma B_{z,\mathrm {osc.}}}{2\pi f_\mathrm{AC}}\right) J_5\left(\frac{\gamma B_{z,\mathrm {osc.}}}{2\pi f_\mathrm{AC}}\right) + J_4\left(\frac{\gamma B_{z,\mathrm {osc.}}}{2\pi f_\mathrm{AC}}\right) J_5\left(\frac{\gamma B_{z,\mathrm {osc.}}}{2\pi f_\mathrm{AC}}\right)\\
 \}
 \end{multline}

 \begin{multline}
 I(-f_{\mathrm{AC}})=i \sqrt{2 \pi }  e^{-5i\tau 2\pi f_\mathrm{AC}}  \{ \\
 - e^{4i\tau 2\pi f_\mathrm{AC}} \left( -1 + e^{i\tau 2\pi f_\mathrm{AC}} \right) J_0\left(\frac{\gamma B_{z,\mathrm {osc.}}}{2\pi f_\mathrm{AC}}\right) J_1\left(\frac{\gamma B_{z,\mathrm {osc.}}}{2\pi f_\mathrm{AC}}\right) \\
 - e^{3i\tau 2\pi f_\mathrm{AC}} \left( -1 +  e^{3i\tau 2\pi f_\mathrm{AC}}\right) J_1\left(\frac{\gamma B_{z,\mathrm {osc.}}}{2\pi f_\mathrm{AC}}\right) \\
 +e^{2i\tau 2\pi f_\mathrm{AC}}  J_2\left(\frac{\gamma B_{z,\mathrm {osc.}}}{2\pi f_\mathrm{AC}}\right)  J_3\left(\frac{\gamma B_{z,\mathrm {osc.}}}{2\pi f_\mathrm{AC}}\right) - e^{7i\tau 2\pi f_\mathrm{AC}} J_2\left(\frac{\gamma B_{z,\mathrm {osc.}}}{2\pi f_\mathrm{AC}}\right) J_3\left(\frac{\gamma B_{z,\mathrm {osc.}}}{2\pi f_\mathrm{AC}}\right)\\
 +e^{i\tau 2\pi f_\mathrm{AC}} J_3\left(\frac{\gamma B_{z,\mathrm {osc.}}}{2\pi f_\mathrm{AC}}\right) J_4\left(\frac{\gamma B_{z,\mathrm {osc.}}}{2\pi f_\mathrm{AC}}\right) - e^{8i\tau 2\pi f_\mathrm{AC}} J_3\left(\frac{\gamma B_{z,\mathrm {osc.}}}{2\pi f_\mathrm{AC}}\right) J_4\left(\frac{\gamma B_{z,\mathrm {osc.}}}{2\pi f_\mathrm{AC}}\right) \\
 -e^{9i\tau 2\pi f_\mathrm{AC}}  J_4\left(\frac{\gamma B_{z,\mathrm {osc.}}}{2\pi f_\mathrm{AC}}\right) J_5\left(\frac{\gamma B_{z,\mathrm {osc.}}}{2\pi f_\mathrm{AC}}\right) + J_4\left(\frac{\gamma B_{z,\mathrm {osc.}}}{2\pi f_\mathrm{AC}}\right) J_5\left(\frac{\gamma B_{z,\mathrm {osc.}}}{2\pi f_\mathrm{AC}}\right)\\
 \}
 \end{multline}




\bibliographystyle{elsarticle-num-names}
\bibliography{Lit}








\end{document}


\section*{Supplementary material}

	\begin{figure}[h]
		\centering
		\includegraphics[width=0.75\linewidth]{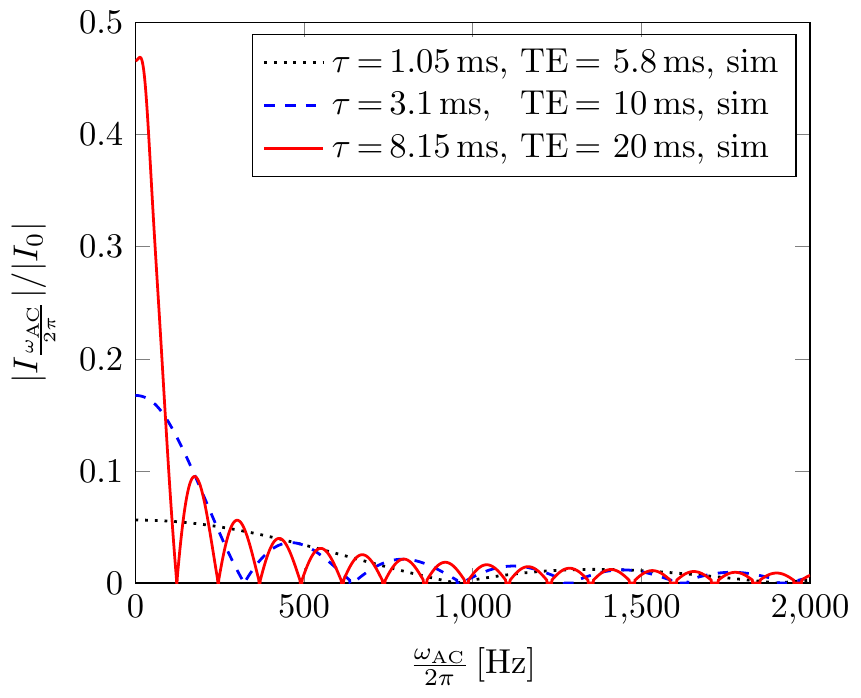}
		\caption{Simulation of the relative signal intensity at different AC frequencies and echo times. The corresponding encoding times are 1.05\,ms, 3.1\,ms and 8.15\,ms. The echo time is related to the encoding time by $\tau = \frac{\mathrm{TE}}{2}-1.85\,\mathrm{ms}$, i.e. only the first echo period reduced by pulses and delays.}
		\label{fig:simACTE}
	\end{figure}
	\begin{figure}[h]
		\centering
		\includegraphics{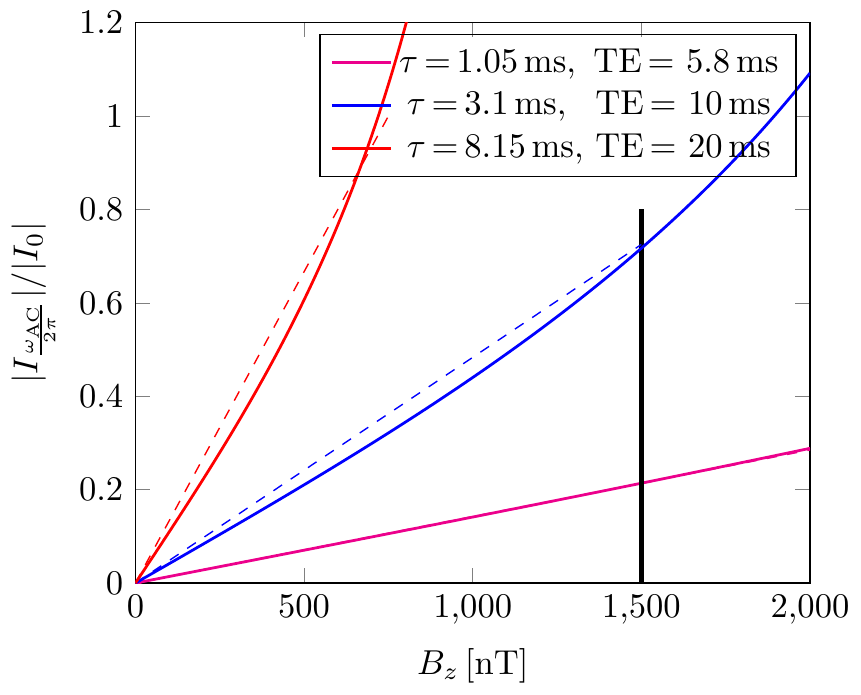}
		\caption{Field dependence of the encoded relative intensity with 1\,Hz AC frequency and 10\,ms echo time. At the fields in this study (gray area), the dependence is approximately linear for the employed echo time of 10\,ms (dashed line). With increasing echo time, the linear region is reduced as can be seen for 20\,ms. }
		\label{fig:fielddep}
	\end{figure}
	\begin{figure}[h]
		\centering
		\includegraphics[width=1\textwidth]{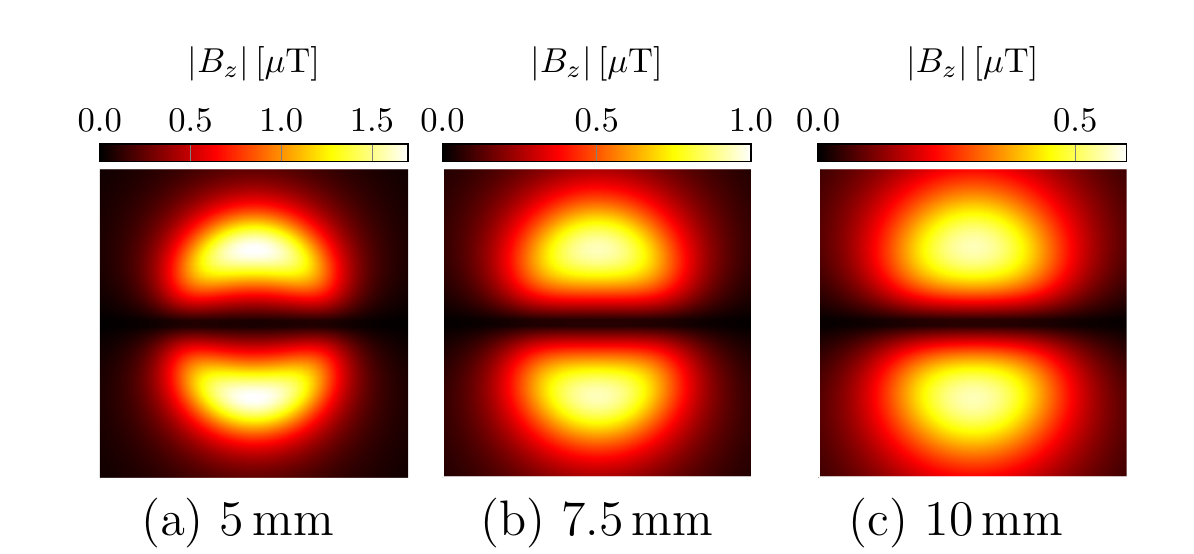}
		\caption{Matlab simulation of a 18\,mm diameter coil with 54\,mA current at three different slice positions.}
		\label{fig:sim}
	\end{figure}
	
	\begin{figure}[h]
		\centering
		\includegraphics[width=1\linewidth]{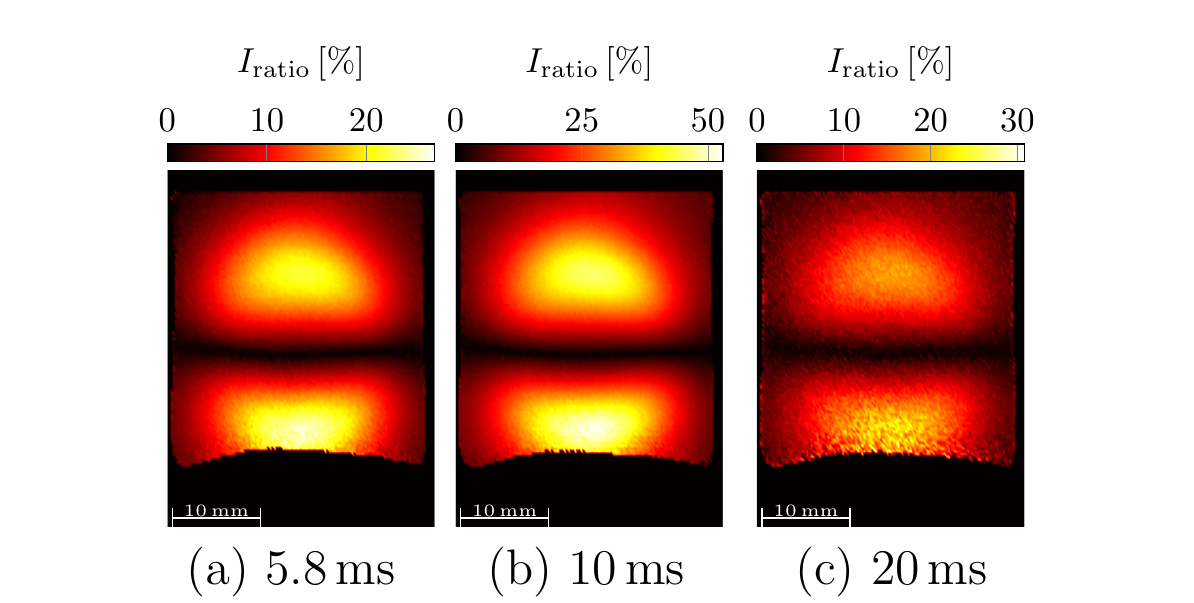}
		\caption{Comparison of the positive frequency part of the spectrum at different echo times and 140\,Hz AC frequency with 54\,mA AC current. While the encoding is stronger at higher echo times, the node pattern results in a lower intensity at 20\,ms. The corresponding encoding times are 1.05\,ms ($t_\mathrm{Echo}$=5.8\,ms), 3.1\,ms ($t_\mathrm{Echo}$=10\,ms) and 8.15\,ms ($t_\mathrm{Echo}$=20\,ms).}
		\label{fig:CompTE}
	\end{figure}

	\begin{figure}[h]
		\centering
		\includegraphics[width=\linewidth]    {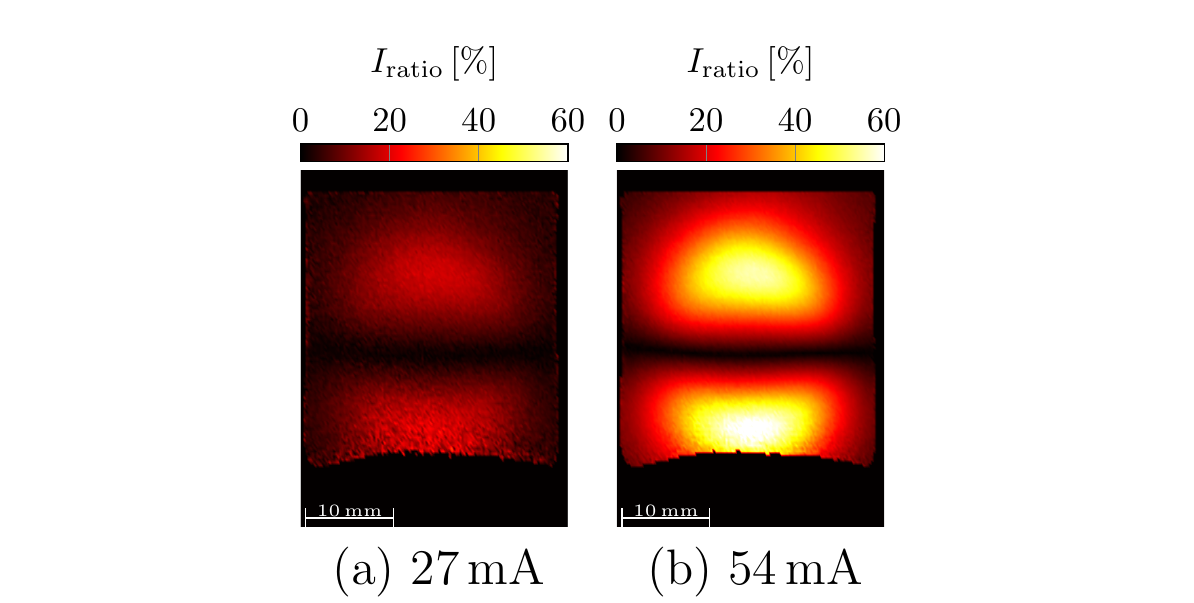}
		\caption{Comparison of different amplitudes of AC currents at 100\,Hz in a coil phantom. 16 experiments with different parts of the waveform have been employed for reconstruction of each point.}
		\label{fig:compcurrent}
	\end{figure}
	\begin{figure}[h]
		\centering
		\includegraphics[width=0.75\linewidth]{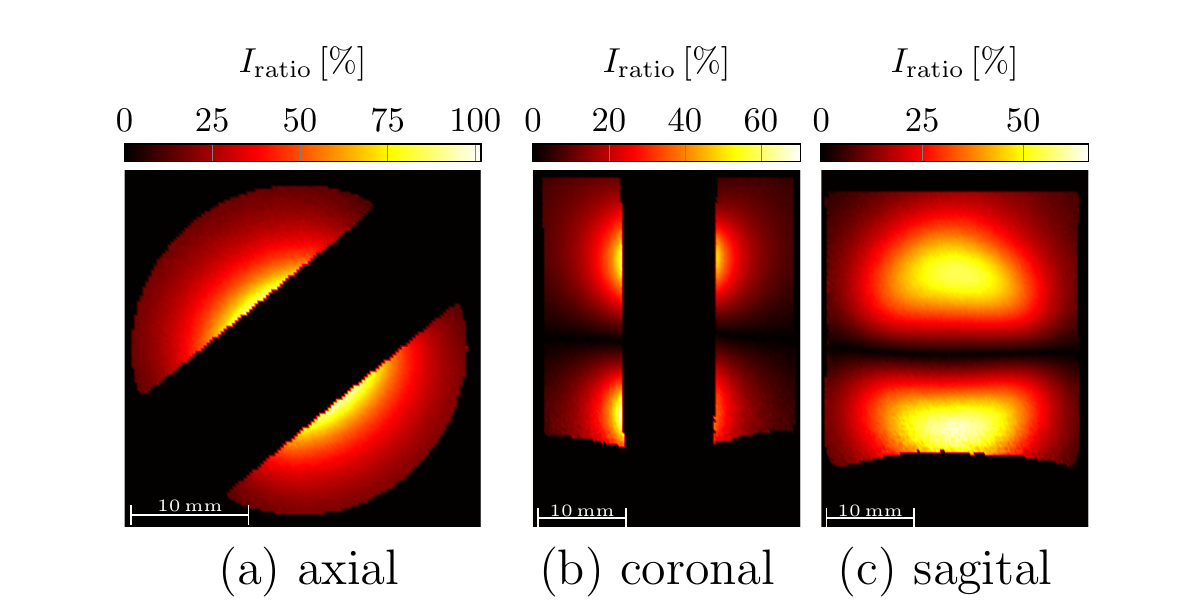}
		\caption{Images at 100\,Hz 54\,mA AC with different orientations. Left: parallel to the coil; Middle: Perpendicular to the coil; Right: Axial slice located in the top. 16 experiments with different parts of the waveform have been employed for reconstruction of each point.}
		\label{fig:3or}
	\end{figure}
	
	\begin{figure}[h]
		\centering
		\includegraphics[width=0.75\linewidth]{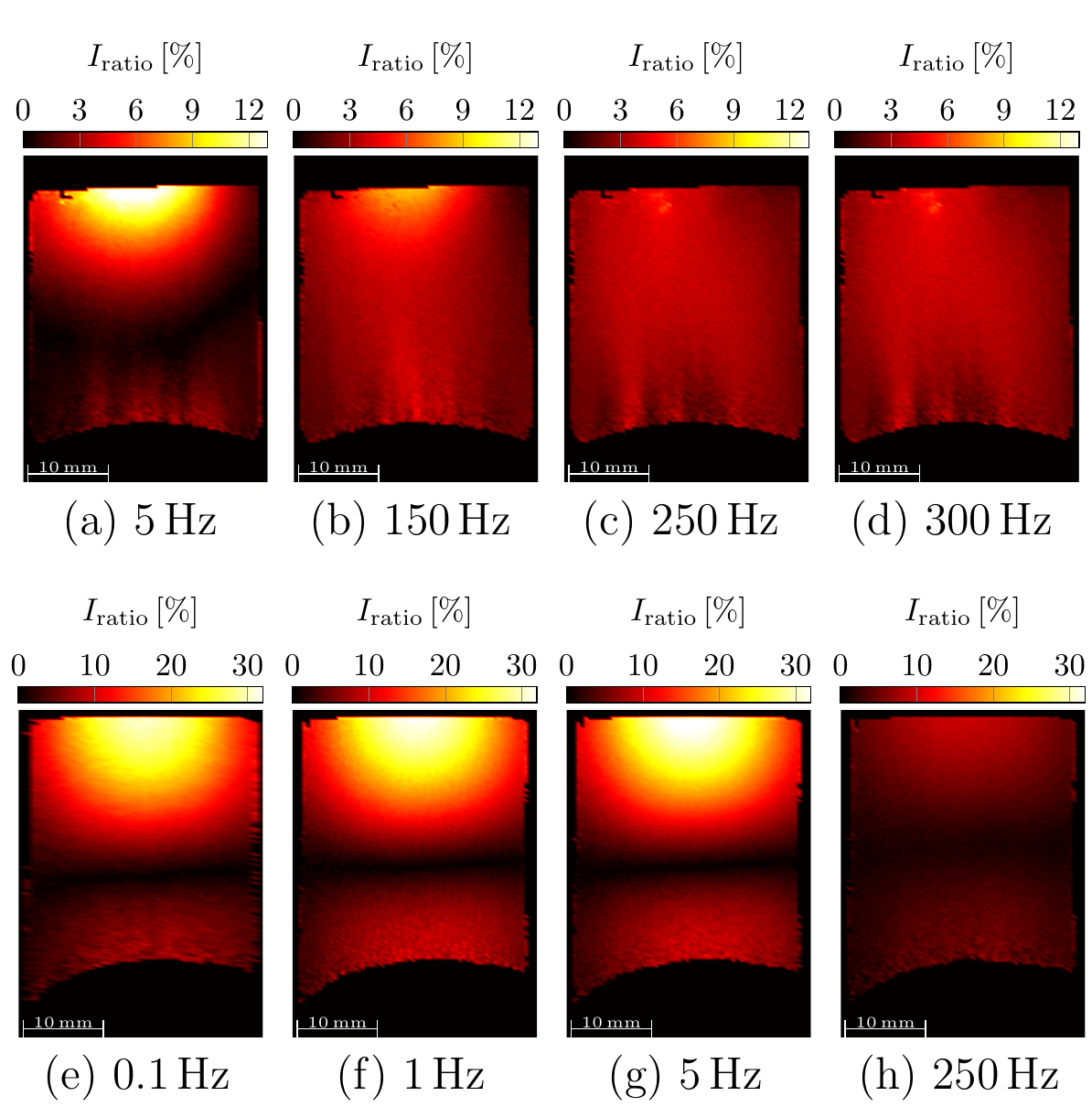}
		\caption{Relative measurements of the oscillating field generated by 50\,mA AC current in various Li-ion batteries at different excitation frequencies without changing the scale. (a)-(d) are maps from an 250\,mAh RIT cell, which is a a  stacked cells. The images in (e)-(h) are from a 600\,mAh PS jelly rolled cell. 16 experiments with different parts of the waveform have been employed for reconstruction of each point of the PS images, while 32 were used for the RIT cell.}
		\label{fig:BattACfreq_unscaled}
	\end{figure}